\documentclass{iaus}
\usepackage{graphicx}

\title[asteroseismic constraints on opacity]
{Asteroseismic constraints on the OPAL opacity interpolation}

\author[Yang, W. M. \& Li, M.]   
{W. M. Yang \and M. Li} \affiliation{Department of Physics and
Chemistry, Henan Polytechnic University, Jiaozuo 454000, China
\break email: yang.wuming@yahoo.com.cn}

\pubyear{2008}
\volume{252}  
\pagerange{ -- }
\date{?? and in revised form ??}
 \jname{The art of modelling stars in the 21st
century} \editors{Licai Deng, K.L. Chan \& C. Chiosi, eds.}
\begin{document}

\maketitle

\begin{abstract}
The frequency difference between a model used only two-point
interpolation of opacity and a model used piecewise linear
interpolation of opacity is of the order of several microHertz at a
certain stage, which is almost 10 times worse than the observational
precision of p-modes of solar-like stars. Therefore, the two-point
interpolation of opacity is unsuitable in modelling of solar-like
stars with element diffusion. 
\end{abstract}

\section{Introduction}
Timescales of evolution and element diffusion are similar in
solar-like stars (\cite[Turcotte et al. 1998]{turc98}). Thus element
diffusion should be calculated in modelling of solar-like stars.
Furthermore, helioseismology has confirmed the importance of
including element diffusion and settling in solar modelling
(\cite[Christensen-Dalsgaard et al. 1993]{chri93}; \cite[Basu et al.
2000]{basu00}).

Heavy-element abundance, $Z$, is constant in main-sequence (MS)
stage of solar-like stars without considering metal settling, thus
evolution of these stars need only one set of opacity tables at a
fixed $Z_{1}$. However the $Z$ is a variable in the models with
metal settling and then the second set of opacity tables must be
obtained at a fixed $Z_{2}$. Opacity at the desired $Z$, $X$, $T$,
and $\rho$ can be obtained by two-point interpolation, i.e.,
\begin{equation}
\kappa(Z, X, T, \rho)=\kappa_{2}(Z_{2}, X, T,
\rho)+\frac{\kappa_{1}(Z_{1}, X, T, \rho)-\kappa_{2}(Z_{2}, X, T,
\rho)}{Z_{1}-Z_{2}}\cdot(Z-Z_{2}). \label{equa1}
\end{equation}

\section{Calculation and results}

In order to study the impact of opacity interpolation, using the
Yale Rotating Evolution Code (YREC7) in its nonrotating
configuration, we construct four models listed in Table \ref{tab1}.
All parameters of the models are same except the parameters of
opacity interpolation. All models evolve from pre-main sequence
(PMS) to somewhere near the end of the MS. The OPAL eos
(\cite[Rogers \& Nayfonov 2002]{roge02}), OPAL opacity
(\cite[Iglesias \& Rogers 1996]{igle96}), and the \cite{alex94}
opacity for low temperature are used. Element diffusion is
implemented following the prescription of \cite{thou94}.

\begin{table}
\begin{center}
\caption{Model parameters. \label{tab1}} {\scriptsize
\begin{tabular}{ccccc}
\hline

Model &  Mass& $Z_{0}$ & Two-point interpolation&
Piecewise linear interpolation \\
      &$M_{\odot}$&  &$Z_{1}$ ----- $Z_{2}$ & $\delta z=Z_{i+1}-Z_{i}$ \\
\hline
MT1 & 1.10 &  0.022 & 0.022 --- 0.021 & ........ \\
MT2 & 1.10 &  0.022 & 0.022 --- 0.020 & ........ \\
MT3 & 1.10 &  0.022 & 0.022 --- 0.019 & ........ \\
MM1 & 1.10 &  0.022 & ....................... & 0.001 \\
\hline
\end{tabular}
}
\end{center}
\vspace{1mm}
 \scriptsize{
 {\it Notes:}
The $Z_{0}$ is the initial metal abundance. Piecewise linear
interpolation: if $Z_{i}\leq Z <Z_{i+1}$, YREC7 will interpolate
between $\kappa(Z_{i})$ and $\kappa(Z_{i+1})$ to obtain the opacity
at the required $Z$. }
\end{table}

\begin{figure}
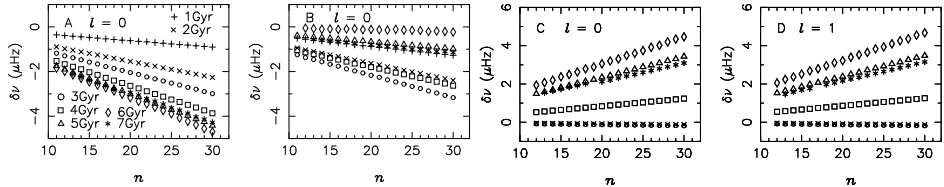

\centering
\includegraphics[angle=-90,scale=0.30]{fig1.ps}
\includegraphics[angle=-90,scale=0.30]{fig2.ps}
\caption{Frequency differences at different ages labeled by the
different symbols. A: $\nu_{MT2}$-$\nu_{MT3}$; B:
$\nu_{MM1}$-$\nu_{MT3}$; C and D: $\nu_{MM1}$-$\nu_{MT2}$.}
\label{fig1}
\end{figure}

The changes in the effective temperature between MT2, MT3, and MM1
at the same age are a few Kelvin, which is within the error of
observation of stellar effective temperature. The frequency
differences between MT1 and MT2 are zero. In Fig. \ref{fig1}A, we
represent the frequency differences between MT2 and MT3 at the same
age. The difference increases from about 1 $\mu Hz$ at the age of 1
Gyr to around 4 $\mu Hz$ at the age of 6 Gyr. But the difference at
the age of 7 Gyr is less than that at the age of 6 Gyr. The
frequency differences between MM1 and MT3 are shown in Fig.
\ref{fig1}B. The differences arrive at a maximum at the age of about
3 Gyr. Then with increase in age, the differences decrease. At the
age of around 6 Gyr, the differences are almost zero. Thus the
discrepancy between MM1 and MT3 mainly occurs between the ages of 2
and 4 Gyr. In Fig. \ref{fig1}C and D, we represent the frequency
differences between MM1 and MT2. The differences are almost zero
when the age of the models is less than 3 Gyr. Then with increase in
age, the frequency differences increase. At the age of around 6 Gyr,
the differences arrive at a maximum. The differences at the age of 4
- 7 Gyr are of the order of several microHertz, which is almost 10
times larger than the uncertainty of observation of stellar p-modes
that is expected to reach 0.1-0.4 $\mu Hz$ (\cite[Th\`{e}ado et al.
2005]{thea05}; \cite[Bedding et al. 2004]{bedd04}).

\section{Discussion}
Heavy-element abundance is a constant in the models of PMS, and the
$Z_{s}$ and $Z_{c}$ are close to the $Z_{0}$ in the early
evolutionary stage of MS. Thus $Z_{1}$ is specified to be $Z_{0}$ in
models MT1, MT2, and MT3, and one of $Z_{i}$ should be equal to
$Z_{0}$ in MM1. The differences of the opacity interpolation in MT2,
MT3, and MM1 should result in the difference between MT2, MT3, and
MM1 and then lead the frequency difference between MT2, MT3, and
MM1. The frequency difference between the models MT2, MT3, and MM1
is of the order of several microHertz at a certain phase, which is
almost 10 times worse than the observational precision of p-modes of
solar-like stars. Consequently, in modelling of solar-like star with
metal settling, the two-point interpolation of opacity is
unsuitable, and at least piecewise linear interpolation is required.

\end{document}